%% file: main.tex
\documentclass[Afour,sagev,times]{sagej}

\usepackage{moreverb,url}

\usepackage[colorlinks,bookmarksopen,bookmarksnumbered,citecolor=red,urlcolor=red]{hyperref}

\usepackage[final, defaultcolor=black]{changes}

\newcommand\BibTeX{{\rmfamily B\kern-.05em \textsc{i\kern-.025em b}\kern-.08em
T\kern-.1667em\lower.7ex\hbox{E}\kern-.125emX}}

\def\volumeyear{2016}

\begin{document}

\def\tCommon{\texttt{Common}}
\def\tCount{\texttt{Count}}
\def\tCompare{\texttt{Compare}}

\def\cTwoDMouse{\emph{2D Vis+Mouse}}
\def\cThreeDMouse{\emph{3D Vis+Mouse}}
\def\cThreeDTrackBall{\emph{3D Vis+Trackball}}
\def\cThreeDVR{\emph{3D Vis+VR}}

\title{Is Embodied Interaction Beneficial? \\ A Study on Navigating Network Visualizations}

\author{Helen H. Huang\affilnum{1}, Hanspeter Pfister\affilnum{1}, and Yalong Yang\affilnum{2}}

\affiliation{\affilnum{1}John A. Paulson School of Engineering and Applied Sciences, Harvard University, USA\\
\affilnum{2}Department of Computer Science, Virginia Tech, USA}

\corrauth{Yalong Yang, Immersion \& Visualization Lab,
Department of Computer Science, Virginia Tech, 
Blacksburg, VA, 24060, USA.}

\email{yalongyang@vt.edu}

\input{body/0-abstract}


\maketitle

\input{body/1-introduction}
\input{body/2-related-work}

\input{body/3-conditions}

\input{body/4-study}

\input{body/5-results}

\input{body/6-discussion}

\input{body/7-conclusion}

\begin{acks}
We would like to thank our reviewers for their valuable comments. We would also like to thank all of our user study participants for their time and feedback. This work was partially supported by NSF grant III-2107328. 
\end{acks}

\bibliographystyle{SageV}
\bibliography{reference}

\end{document}

%% file: body/0-abstract.tex
\begin{abstract}
Network visualizations are commonly used to analyze relationships in various contexts, such as social, biological, and geographical interactions. To efficiently explore a network visualization, the user needs to quickly navigate to different parts of the network and analyze local details. Recent advancements in display and interaction technologies inspire new visions for improved visualization and interaction design. Past research into network design has identified some key benefits to visualizing networks in 3D versus 2D. However, little work has been done to study the impact of varying levels of embodied interaction on network analysis. We present a controlled user study that compared four network visualization environments featuring conditions and hardware that leveraged different amounts of embodiment and visual perception \added{ranging from a 2D visualization desktop environment with a standard mouse to a 3D visualization virtual reality environment}. We measured the accuracy, speed, perceived workload, and preferences of 20 participants as they completed three network analytic tasks, each of which required unique navigation and \added{substantial effort to complete}. For the task that required participants to iterate over the entire visualization rather than focus on a specific area, we found that participants were more accurate using a VR HMD and a trackball mouse than conventional desktop settings. From a workload perspective, VR was generally considered the least mentally demanding and least frustrating to use in two of \added{our} three tasks. It was also preferred and ranked as the most effective and visually appealing condition overall. However, using VR to compare two side-by-side networks was difficult, and it was similar to or slower than other conditions in two of the three tasks. Overall, the accuracy and workload advantages of conditions with greater embodiment in specific tasks suggest promising opportunities to create more effective \added{environments in which to analyze network visualizations.}
\end{abstract}

%% file: body/1-introduction.tex
\section{Introduction}

The ability to navigate to different parts of a visualization is an essential component of visual data analytics~\cite{kerren_visual_2008}.
Navigation techniques for visualizations have been extensively studied in the visualization and human-computer interaction communities, mainly for visualizations on flat 2D screens~\cite{Cockburn09}.
In those traditional computing setups, most analysts use a mouse as an input device to pan around the visualization and zoom in and out to check details.
More recent display and interaction devices allow us to navigate visualizations in more interactive and embodied ways. 
\added{These newer combinations of displays and devices can} bring potential benefits such as reduced cognitive load \added{because they enable} greater direct navigation and reorientation of the visualization ~\cite{dourish_where_2004}\added{, which we will refer to as graph ``manipulation"}.

There is growing interest in using immersive environments (i.e., virtual and augmented reality, or VR/AR) for data visualization~\cite{Marai16,ens_grand_2021}.
Two motivations are often reported for the use of immersive visualization.
\textbf{First,} we can render stereoscopic 3D visualizations in VR/AR. While some data visualizations such as pie charts are negatively impacted using 3D, others such as network visualizations see advantages including reduced visual clutter~\cite{Ware94,Ware05,Ware08,Kwon16}. 
\textbf{Second,} we can perform direct, embodied manipulation in VR/AR. 
When using \added{the typical} mouse and 2D screen, the physical interaction space is separated from the digital display space, meaning a user must manipulate a tangible, physical device to interact with the intangible digital graphics on the screen.
However, in VR/AR, the interaction and display space are the same physical space, potentially reducing cognitive load by removing the cost of context-switching between physical and digital spaces. 

\begin{figure*}
  \centering
  \includegraphics[width=\textwidth]{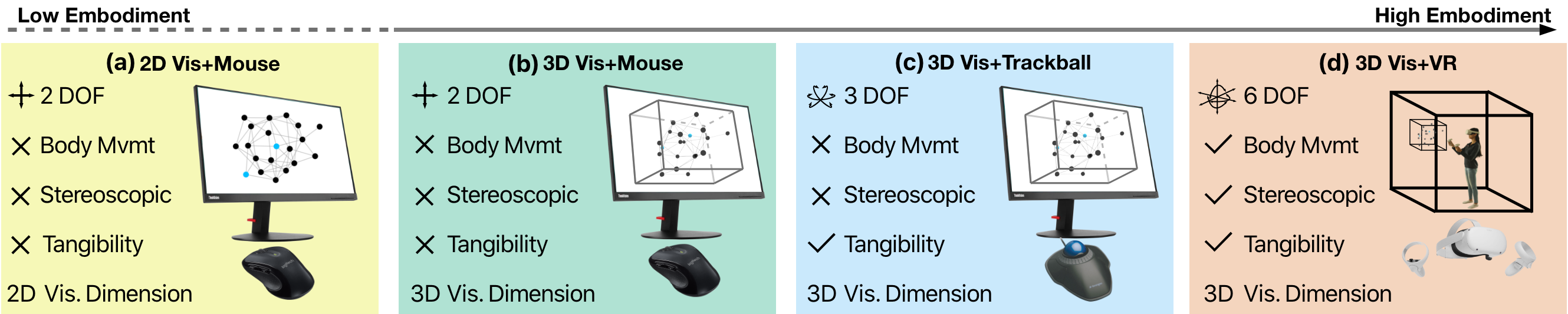}
  \caption{Tested conditions in our user study with key characteristics: (a) 2D network visualization displayed on a 2D screen with a standard mouse, (b) 3D network visualization on a 2D screen with a standard mouse, (c) 3D network visualization on a 2D screen with a trackball mouse, (d) 3D network visualization in virtual reality with hand-held controllers.}
  \label{fig:condition}
\end{figure*}

The benefits of using 3D visualizations \added{over 2D ones} have been \added{well} studied, with existing research finding that scatterplots~\cite{kraus_impact_2020,Bach18}, network visualizations~\cite{Kwon16}, and some geographic visualizations~\cite{yang_tilt_2020,filho_evaluating_2020} are more effective in 3D than in 2D. 
Conversely, the benefit of direct manipulation and greater embodiment in VR/AR has been less explored.
There are two studies that are most relevant to our work.
In one study, Bach et al.~\cite{Bach18} \added{also compared a spectrum of experimental conditions} including desktop, tablet AR, and HoloLense conditions, with a focus on 3D scatterplots \added{rather than networks}. However, their tasks only involved mark-based interactions, which are less intuitive than embodied interactions. 
In another study, Kraus et al.~\cite{kraus_impact_2020} again compared 3D scatterplots on a 2D screen to those in VR environments with different sizes. However, participants could only move around the visualization in VR and were not able to manipulate the visualizations in an embodied way. 
While physical movement is a way to navigate around a visualization in an immersive environment, it introduces more physical workload and is restrained by the physical space available~\cite{lages_move_2018}, \added{so relying solely on movement to navigate a visualization may not be ideal}. 
Both of these studies lacked the element of embodied interaction, which allows for direct manipulation of objects and provides a more intuitive, life-like experience while reducing the physical movement involved in analyzing a visualization.

To fill this gap, we studied \textbf{the effect of embodied interaction through different display and interaction devices on navigating network visualizations.}
We were interested in how different commercialized devices offer graph visualization manipulation and how their different capabilities influence people's performance in visualization navigation. 
Specifically, we compared four conditions in a controlled user study (see \autoref{fig:condition}): 2D and 3D network visualizations with a standard mouse, 3D network visualization with a trackball mouse, and 3D visualizations with a VR head-mounted display (HMD) and corresponding controllers.
While there was little embodiment difference between the two conditions using a standard mouse, the trackball mouse had more embodiment than a standard mouse because the physical trackball acted as a tangible \added{direct} proxy for the 3D visualizations. 
Rotating the trackball provided a closer match between the 3D interaction and the \added{3D} display space.
\added{The} VR environment had even more embodiment, as users could directly manipulate visualizations using 6D tracked controllers (i.e., 3D position and 3D rotation) similarly to manipulating real-life objects.
We chose network visualizations as the data visualization to study because they require a significant amount of navigation effort in many analytic tasks and have been less studied with different interaction and display devices.

\added{In our study} we asked 20 participants to complete the following three fundamental analytic tasks we derived from \added{related work}~\cite{Ware94,Ware05,Ware08,Kwon16,cordeil_immersive_2017,Drogemuller20,Kotlarek20} and widely-accepted visualization task taxonomies~\cite{Bach18,kraus_impact_2020,besancon_state_2021,hand_survey_1997}: (1) identify the \emph{common} nodes between two nodes \added{in a visualization} (\tCommon{}), (2) \emph{count} the number of triangles \added{in a visualization} (\tCount{}), and (3) \emph{compare} two network visualizations (\tCompare{}). 
In each task, we measured the accuracy, speed, perceived workload, and preferences of the participant, and we found that using a trackball mouse and VR HMD was more accurate in \tCount{} but required more time than a standard mouse in \tCommon{}. 
In addition, participants struggled in VR when comparing two side-by-side network visualizations, resulting in long completion times and low accuracy.
From a workload perspective, participants found 3D network visualization to be less mentally demanding and frustrating when compared to 2D alternatives.
Participants found the trackball mouse to be more mentally and physically demanding than a standard mouse, and VR was generally considered to be the least mentally demanding and frustrating for \tCommon{} and \tCount{}, though the opposite was true for \tCompare{}.
After collecting participant preferences on the aesthetics and effectiveness of the different conditions, we found that participants ranked the typical desktop set up with a standard mouse and 2D network visualization to be the \textit{least} visually appealing and effective, while participants ranked VR to be the overall most effective and aesthetically pleasing condition. 
\added{The combination of all our} results suggests that design and interaction improvements can be made to the way modern network visualization analysis is conducted \added{to help analysts better navigate network diagrams}. 

%% file: body/2-related-work.tex
\section{Related Work}
\label{sec:related-work}

\noindent\textbf{Visualization navigation.}
To efficiently explore a visualization, the user must be able to easily navigate to different parts of the visualization and inspect local details.
Several interaction approaches such as focus+context, overview+detail, and pan\&zoom have emerged to support visualization navigation, and they have been extensively studied for 2D visualizations~\cite{Cockburn09,tominski_survey_2014,tominski_interactive_2017}.
Conversely, interaction approaches with more immersive visualizations have been less widely explored, though some work in this area has indeed been done. Yang et al.~\cite{yang_embodied_2021} compared overview+detail and pan\&zoom interfaces for 3D scatterplots in VR.
Although they did not come to a conclusion on which condition performed better, they did find that participants preferred the pan\&zoom interface.
Similarly, Drogemuller et al.~\cite{Drogemuller20} compared three locomotion methods and the overview+detail (or Worlds-In-Miniature) techniques for network visualizations.
However, they only considered room-sized visualizations in VR, which are uncommon in current visual analytics workflows. 
Lages and Bowman~\cite{lages_move_2018} studied more realistic analytics conditions by comparing the effectiveness between manipulating a visualization and moving around a visualization, and they found that some participants performed better by manipulating \added{the graph view} than physically moving around the graph.
Walking is one of several primary ways that 3D User Interfaces (3DUI) and Virtual Reality have allowed users to navigate and move through 3D immersive space, as classified by Laviola et al.~\cite{laviola20173d}.

We chose to use the pan\&zoom manipulation technique for two main reasons. 
First, it is a more widely \added{used} navigation method and is seen in software such as Google Maps and digital image viewers, lowering the barrier to using it in our study. 
Second, the pan\&zoom technique can be similarly replicated across the different devices used in our study. 
Using a more intuitive and familiar navigation technique allowed participants' performance to be more directly impacted by the level of embodiment provided by the devices in our four conditions, rather than \added{by any} difficulty with understanding a new navigation method.

\smallskip
\noindent\textbf{2D, 3D, and immersive network visualizations.} 
Node-link diagrams are the most common and intuitive methods of displaying network visualizations.
Researchers have extensively studied different layout algorithms for creating node-link diagrams~\cite{diaz_survey_2002,tamassia2013handbook}, and force-directed layouts are one of the most widely used methods due to their simple implementation and \added{their mitigation of link crossings within a network.}
As a result, we used a force-directed layout to produce the node-link diagrams for our user study.
\added{Despite the mitigated link crossings in a force-directed layout,} 2D node-link diagrams in a limited display space \added{such as a computer screen can result in visual clutter} that makes it difficult for the viewer to perceive information effectively~\cite{Ware08}.

\added{On the other hand, with one extra dimension, visualizing networks in 3D has the potential to address or at least alleviate this issue.
A series of user studies has confirmed the advantages of 3D over 2D displays in displaying node-link diagrams, with the most representative studies being from Ware et al.~\cite{Ware94,Ware05,Ware08}, Greffard et al.~\cite{greffard2011visual,greffard2014beyond} and Alper et al.~\cite{alper2011stereoscopic}.}
However, due to the occlusion and perspective distortion found in 3D visualizations, it is essential that users can easily change their viewing position and direction \added{while} intuitively manipulate the digital graph view\cite{Ware08,ens_grand_2021}.
In VR, Kwon et al.~\cite{Kwon16} implemented an egocentric layout to show networks with clusters. 
\added{With this layout,} they found network visualizations performed better in VR than on a 2D screen, especially with difficult tasks.
Cordeil et al.~\cite{Cordeil17} compared network visualization in VR and on a CAVE screen for collaborative analysis, \added{where} they found VR had advantages in completion time.
Kotlarek et al.~\cite{Kotlarek20} compared 3D immersive network visualizations with their 2D desktop alternatives and found that VR contributed to better interpretation of the network structure, while 2D resulted in better spatial memory.
However, in those studies, participants could not manipulate their views, which is considered an important feature for immersive visualization. Those studies also only considered the comparison between a limited number of devices.
\added{Kraus et al.~\cite{kraus2022immersive} provided a comprehensive review of visualizations in immersive environments. Specifically, as they pointed out, there is a large design space to explore for immersive network visualizations.}
To expand on all this existing research, we designed our study to focus on \added{the effect of embodied view manipulation on graph navigation}, and we compared four \added{computing environments} that cover a wider range of the embodiment spectrum than devices used in other studies.

\smallskip
\noindent\textbf{3D interactions with a standard mouse.} 
Interacting with 2D visualizations has traditionally involved using a mouse to point, select, and manipulate the view.
\added{However, when interacting with 3D visualizations,} using a standard mouse poses an issue due to the mouse's limited DOF.
Chen et al.~\cite{chen_study_1988} \added{used a standard mouse} to compare different 3D interaction methods and found that the \emph{``virtual sphere''} method performed best. 
It simulates a \cThreeDTrackBall{} by encasing the 3D view in an invisible sphere that a standard mouse can then drag to rotate.
The ``virtual sphere'' method (also known as orbit control) has been widely used in commercial applications such as CAD and was \added{thus used} in our study (\autoref{fig:condition}(b)).
\added{Of note, though, is that} extra mental effort is expected to map \cTwoDMouse{} movements to a 3D virtual sphere due to the inconsistency in DOF between the input device and digital object.

\smallskip
\noindent\textbf{Tangible proxies for 3D interaction.} 
Tangible 3D input devices can reduce the additional mental effort of a \cTwoDMouse{} and better facilitate 3D interaction than \cThreeDMouse{}.  
Various studies have confirmed the benefits of using 3D input devices \added{as physical proxies}~\cite{feick_tangi_2020,Dinh99,Hoffman98,reinhardt_impact_2018,Besancon17,Englmeier19,englmeier_tangible_2020}.
Meanwhile, Hand~\cite{hand_survey_1997} and Besançon et al.~\cite{besancon_state_2021} provided comprehensive reviews of such 3D devices.
However, many tangible 3D input devices (e.g., those ones in~\cite{Englmeier19,englmeier_tangible_2020,satriadi2022tangible}) are customized for studies and not easily accessible to ordinary users.
Thus, in our study, we chose to test a broadly available trackball mouse as our proxy.
While the interaction is similar to that of the ``virtual sphere'', the trackball mouse differs \added{from a standard mouse} by providing a tangible 3D ball (instead of a virtual ball) that rotates the 3D object on the screen exactly how the physical trackball in a user's hand rotates (\autoref{fig:condition}(c)).
The \cThreeDTrackBall{} condition increases embodiment over the \cThreeDMouse{} condition by directly mapping 3D interactions to 3D views.
However, the interaction space (i.e., the desk's surface) is \added{still} spatially separated from the display space (i.e., the screen), which introduces some extra cognitive load.

\smallskip
\noindent\textbf{Immersive VR interactions.} 
Commercial VR head-mounted displays (HMDs) provide a fully immersive experience at an affordable cost. They allow the user to see stereoscopic 3D views and to manipulate them with 6-DOF hand-held controllers (see \autoref{fig:condition}(d)), which most closely follows the \emph{``what you do is what happens''} paradigm~\cite{Bowman08}.
\added{Specific to immersive node-link diagrams, Sorger et al. explored interactions to study egocentric views of network visualizations~\cite{sorger2019immersive,sorger2021egocentric} and Drogemuller et al.~\cite{Drogemuller20} studied different locomotion methods for navigating immersive network visualizations.
However, their studies did not directly compare immersive environments to other computing environments.}
Meanwhile, some studies have found benefits to using VR over the conventional \cTwoDMouse{} display environments with different visualizations such as scatterplots~\cite{kraus_impact_2020,Bach18,wagner_filho_virtualdesk_2018}, network visualization~\cite{Kwon16,Huang17}, space-time cubes~\cite{filho_evaluating_2020}, and visual channels~\cite{Whitlock20}.
As far as we know, though, no study has compared VR to a condition with a 3D input device on a 2D display (our \cThreeDTrackBall{} condition) for network visualizations.

%% file: body/3-conditions.tex
\section{Rationale And Experimental Conditions}
\label{sec:technique}
Initially, the use of 3D for visualizations was not commonly appreciated in the literature~\cite{munzner2014visualization}. 
However, a series of more recent studies have provided empirical evidence for the benefits of 3D over 2D, especially with improving display and interaction technologies.
As pointed out by Marriott et al~\cite{immersive_analytics_immersive_2018}: \textit{``it is time to reconsider the value of 3D for information visualization.''}
\added{So,} based on previous work, we used 3D network visualizations in three of our testing conditions, namely 3D Vis with a mouse \added{(\cThreeDMouse{})}, a trackball mouse \added{(\cThreeDTrackBall{})}, and VR \added{(\cThreeDVR{})}. 
By comparing these three different environments, we aimed to investigate the effect of different levels of embodiment \added{(in terms of display and device)} on navigating a network visualization.
Notably, we also included a condition using 2D network visualization to both confirm the benefits of using 3D visualizations over 2D ones in our tasks and enrich empirical knowledge of the comparisons between 2D and 3D network visualizations.

To systematically investigate embodiment, we reviewed previous studies~\cite{Ware94,Ware05,Ware08,Kwon16,cordeil_immersive_2017} and taxonomies on visualization interaction~\cite{Bach18,kraus_impact_2020,besancon_state_2021,hand_survey_1997}, and identified five fine-grained properties of embodiment:

\textbf{\textit{Visualization Dimension}}---whether the network visualizations are rendered in 2D or 3D.

\textbf{\textit{DOF}} (i.e., Degree of Freedom)---the extent of the input device's rotational and translational freedom of movement.

\textbf{\textit{Body Movement}}---whether the user could leverage body movement to change view point and direction.

\textbf{\textit{Stereoscopy}}---whether the display enabled a stable depth perception of head-tracking stereoscopic visuals.

\textbf{\textit{Tangibility}}---whether the user could tangibly interact with the visualization \added{either through a physical proxy of the visualization or a 3D virtual representation of it \textit{as if} touching the visualization itself. Tangibility using a proxy would be considered a weaker form of tangibility \added{than tangibility using virtual reality controllers to manipulate a graph view.}}

Testing the effect of every single property was not feasible, as it would result in too many conditions for participants in the study. \added{However, using these five properties, we imagined a general spectrum of environments with varying levels of embodiment based on the number of properties they encapsulated. Environments with more limited DOF, less body movement, no stereoscopy, less tangibility, and only 2D visualizations were considered to give the user low embodiment whereas environments that allowed for more DOF, greater body movement involvement, stereoscopy, more tangibility, and that featured 3D visualizations were considered more embodied.}
\added{To most reasonably decide on the environments along the spectrum we would use for the study, we decided to follow} Bach et al.'s strategy~\cite{Bach18} and chose easily accessible hardware environments \added{with different numbers of the five properties and therefore different levels of embodiment} ( \autoref{fig:condition}). 
\added{One of our key goals for this project was to conduct research that could be applicable to the general public, which is why we emphasized the importance of choosing accessible hardware devices even if they were located at intervals along the spectrum of environments that were not exactly equal.}

\added{The following were our four computing environments:}

\textbf{\cTwoDMouse{}}: used a standard mouse with 2 DOF as the input device and a 2D monitor to render the visualizations. The network visualizations were rendered in 2D and participants used the mouse to pan\&zoom with the views. 
To pan, the user could left-click and drag the visualization around the screen. To zoom, they would use the scroll wheel on the mouse.

\textbf{\cThreeDMouse{}}: used the same setup as \cTwoDMouse{} but rendered 3D network visualizations rather than 2D diagrams and used the aforementioned ``virtual sphere'' method~\cite{chen_study_1988}. 
Left-click and drag were used to rotate the 3D visualization while right-click and drag would pan the visualization around the screen. 
To zoom, the user would still use the scroll wheel like in \cTwoDMouse{}.

\textbf{\cThreeDTrackBall{}}: used a 2D monitor to render 3D network visualizations. Instead of using a traditional \added{computer} mouse, 
\cThreeDTrackBall{} used a trackball mouse
that was a stationary device with a physical ball that a user rotated with their hand, \added{allowing for 3-DOF}. \cThreeDTrackBall{} featured two modes. Cursor mode was the default mode and allowed users to rotate the physical trackball to move the cursor. Upon a left click, the trackball would enter rotation mode, where rotating the trackball rotated the 3D graph.
To zoom, the user used the trackball mouse's scroll ring.

\textbf{\cThreeDVR{}}: used a head-mounted display (HMD) to render stereoscopic 3D visuals and two hand-held VR controllers to provide direct, 6-DOF manipulation.
\added{Users could use the controllers to grab, move, and reorient the visualizations to find the desired information by pointing both controllers at the visualization, pressing the trigger buttons on the front of the controllers, and moving their hands in the same logical direction.
To zoom,} 
we used the built-in zoom implementation from the MRTK package~\cite{MRTK}, which contains a standard zoom implementation used in many other commercial and open-source platforms, such as SteamVR, Oculus, and VRTK. 
\added{Participants could not only pinch-to-zoom the graph by grabbing two parts of the graph and moving both hands closer or farther apart while holding the triggers, but they could also bring the visualizations closer to themselves.
To make solution selections, participants could use the trigger on one controller to select a node while hovering over it (for \tCommon{} and \tCompare{}).}
\added{In addition to manipulating the graph view,} participants could freely move around the diagram to effectively inspect different parts of the visualization. 

\added{One major difference between this condition and the other conditions on the desktop was the view box. 
While the view boxes of the graphs in the desktop environments occupied almost the entire screen, they were still limited by the size of the screen.
As a result, parts of the graph would not be seen if the user zoomed in closely on other areas of the graph.
This characteristic was much less obvious in the VR environment.
Despite still having graphs rendered in a specific bounding box area in the virtual environment, the VR environment allowed the entire virtual space around the user to essentially become the ``screen". 
Consequently, users could enlarge the graph to be much closer and bigger without ``cutting off" any parts of it within the environment.
However, there was still a view limitation within VR known as the field of view (FOV) of the VR headset, which is a concept that represents the amount of a virtual world a user can see at once. 
For the Oculus Quest 2, the FOV is 89 degrees. 
So, even though the entire graph in VR could be seen if the user moved their head, unlike on a computer screen, only parts of the graph would be visible at once to a user if the graph were too close in the virtual environment. 
We decided not to include explicit view boxes in \cThreeDVR{} like on the computer screen to allow users to experience the natural differences between using VR and desktop environments. 
}

\smallskip\textbf{Other Design Choices.}
We used a standard force-directed layout algorithm implemented in d3js\footnote{https://github.com/d3/d3-force} and its extension\footnote{https://github.com/vasturiano/d3-force-3d} to calculate the node positions for all network visualizations, both in 2D and 3D, given the popularity and wide use of this method.
\added{We used the default option from the algorithms to render the starting perspectives of the 3D node-link diagrams.}
We also used a standard white HTML background for conditions \added{in the desktop environments} and the default MRTK~\cite{MRTK} scene in the virtual reality environment.

\added{In addition, we considered other interesting design areas related to the idea of controllers versus gestures, haptics, and manipulation versus movement, to name a few. 
In the end, our decisions to use controllers, not implement input device haptics, and allow for movement were all based on precedent work we studied, a desire to create the most seamless user experience (e.g. VR HMDs are more sensitive to controller actions than gestures), and our goal of using conditions that are generally more accessible to the general public (e.g. devices using haptics are not as widely available as commercial HMDs without haptics).}

%% file: body/4-study.tex
\section{User Study Design}
In this section, we explain the details of our setup, participants, procedure, tasks, data generation, measures, and hypotheses of our controlled user study.

\begin{figure*}
  \centering
  \includegraphics[width=\textwidth]{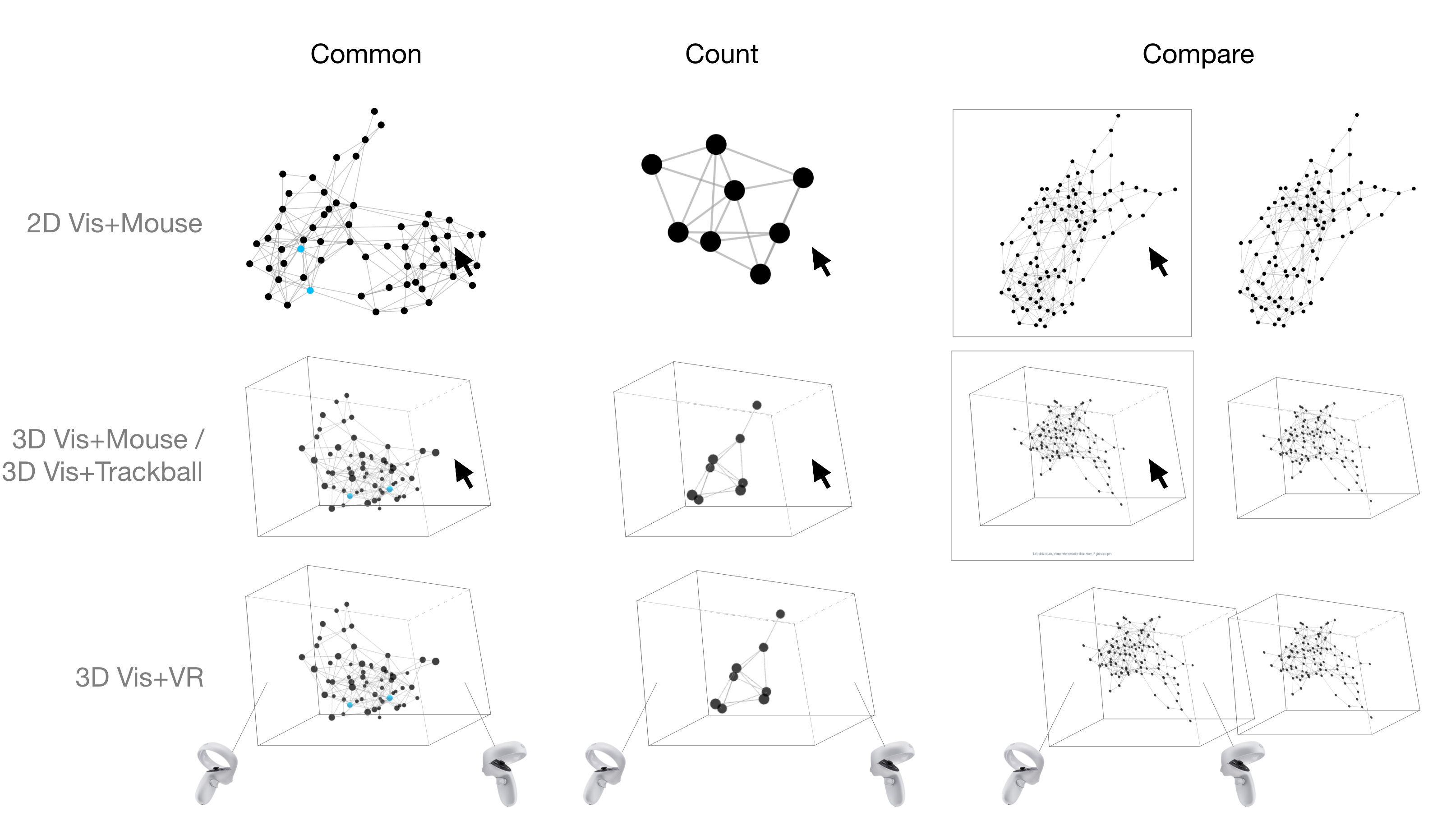}
  \caption{Users performed three different graph tasks in our study. (a) Find the common nodes between two highlighted nodes, (b) count the number of triangles in a graph, (c) find the missing nodes between two side-by-side graphs. Above are examples of 3D graphs used in each task.}
  \label{fig:tasks}
\end{figure*}

\subsection{Experimental Setup}
\cTwoDMouse{}, \cThreeDMouse{}, and \cThreeDTrackBall{} used a 23.8'' 2D flat screen with a resolution of 2560$\times$1440. 
\cTwoDMouse{} and \cThreeDMouse{} used a standard mouse.
\cThreeDTrackBall{} used a Kensington orbit trackball mouse.
\cThreeDVR{} used an Oculus Quest 2 HMD with a resolution of 1832$\times$1920 per eye paired with its two hand-held controllers as input devices.
To reduce the complexity and avoid the potential confounding effect, sensitivity settings were fixed for all devices. Participants were not given the option to customize input device sensitivity (e.g., mouse speed).

\subsection{Participants}
We recruited 20 participants, 12 male and eight female, through university mailing lists to complete the study. All 20 were undergraduate students from a wide range of STEM and non-STEM \added{majors}, with 6 participants \added{majoring} in Computer Science. All participants were between 18 and 24 years old, and all had either normal or corrected-to-normal vision.
Regarding experience with a mouse, all but one participant noted significant experience. The one participant without significant mouse experience indicated between 0 and 10 hours of lifetime mouse use and primarily used a laptop touchpad instead. 
Regarding trackball experience, 16 participants had never used a trackball before, two had used one for fewer than ten hours, and two used a trackball either daily or for at least more than 20 hours in their lifetime. 
Regarding VR experience, four participants had never used a VR HMD before, 15 had used it for fewer than ten hours and sometimes only in the context of basic Google Cardboard devices, and one had used VR between ten and 20 hours.

\subsection{Procedure}
The experiment followed a within-subject design. We used a Latin square design to determine the order of the conditions each user would use to complete the study. This was done to mitigate learning effects.
Each participant completed 24 study trials: 4 conditions $\times$ 3 tasks $\times$ 2 \added{study} trials.
Each participant also completed the same number of training trials to ensure they understood the task and were familiar with the conditions before conducting the study trials.
The study was conducted in-person and took on average two hours per participant. We compensated each participant with a \$20 gift card.

\added{To begin the study}, participants were given a brief introduction to the experiment.
They were then asked to complete the tasks in the order of \tCommon{}, \tCount{}, and \tCompare{}. 
\added{For each task,} participants were first presented two practice trials to ensure that they were familiar with the visualization, interaction, and task before being given their two official study trials. 
When introduced to new conditions, participants completed a short training session that explained and demonstrated the device setup.
After each task, participants filled out a survey adapted from NASA's Task Load Index~\cite{Nasa} to rate the four conditions. 
We also asked participants to provide verbal feedback highlighting the positives and negatives associated with each condition for that task. 
After completing all three tasks, participants completed a post-study survey where they ranked the conditions overall for effectiveness and aesthetics and provided demographic information. 

Conducting in-person user studies and recruiting participants during the pandemic was extremely challenging. We strictly followed COVID-19 policies to ensure participants' and investigators' safety, e.g., we ensured a two-hour gap between sessions and sanitized all devices before and after each session. 

\subsection{Tasks and Data}
To better understand the effect of embodiment, we selected \added{our \tCommon{}, \tCount{}, and \tCompare{}} tasks because they required \added{substantial} interaction effort.
We referenced the network visualization task taxonomy by Lee et al.~\cite{Lee06} and relevant user studies~\cite{Kwon16,Drogemuller20,cordeil_immersive_2017,Kotlarek20,yoghourdjian_exploring_2018} to choose these three representative tasks, which follow the design space of visualization tasks by Schulz et al.~\cite{schulz_design_2013} to cover targets in different levels of detail.

We now describe the details of our three chosen tasks. 
All study stimuli are also included in the supplementary materials.

\textbf{\tCommon{}}: \textit{Find the common node \added{neighbors} between two highlighted nodes}. 
This task investigated the ability of different testing conditions to enable participants to closely explore \emph{a given part of the node-link diagram}. Participants first had to navigate to the part of the network visualization with the two highlighted nodes, then identify the nodes that were linked to \added{(i.e. common to and neighbors with)} both highlighted nodes.
This is a common connection topology task and has been used by Kwon et al.~\cite{Kwon16}. 

\added{In more detail, when participants began each trial for this task, the internal timer (invisible to the participants) would start and participants would be shown a network visualization with two of its nodes highlighted blue.}
Participants needed to find the nodes that were connected to both highlighted nodes and click them.
\added{After clicking on a node, the node would turn red and would be considered part of the participant's answer. 
If participants changed their minds, they could click on the node again to return it to its original color and remove it from their final answer.
When they were sure of their selections, participants would then click a "Done" button on the page which would lock in their selections, stop the internal timer, and move them to the next page}. 

\added{Given the density of nodes, this task required participants to exert effort zooming in on the area with the two highlighted nodes and rotating the graph (when applicable) to identify neighbors and confirm assumptions.}
Each study trial had 60 nodes separated into three even-sized clusters, with two to three common nodes per trial.
The probability that a link existed between two nodes was 0.2 within a cluster and 0.05 between clusters.

\textbf{\tCount{}}: \textit{Count the number of triangles in a network visualization}. 
Users were asked to count all triangles formed by the nodes and links in a visualization, which targeted the importance of finding cliques or strongly connected components in a network. 
This task investigated both navigation and spatial memory capabilities offered by the different testing conditions. 
Drogemuller et al.~\cite{Drogemuller20} and Cordeil et al.~\cite{cordeil_immersive_2017} used this task in their user studies \added{as well}.
\added{Here, participants needed to iterate over \emph{the entire network visualization} and identify each triangle without double counting the same triangle from different angles. 
Once they were certain of the number, they would click an ``input number" button on the page that would stop the internal timer and move them to a new page, which removed the visualization and displayed only a number slider. 
Using the slider, participants would indicate the number of triangles they counted and then click another button to submit their answer, without the time taken to use the slider counted toward their completion time.}

\added{In terms of the effort needed for this task, \tCount{} required participants to exert effort when dynamically rotating the diagram in smooth and logical ways to properly count every triangle in the graph.}
Each study trial had eight nodes in one cluster with either 15 or 17 links between nodes. We also controlled the number of triangles within the range of six to 11.
Our internal test revealed significant mental and physical fatigue for any larger or more complex network for this task.

\textbf{\tCompare{}}: \textit{Find the missing nodes between two side-by-side network visualizations}. 
Participants were given two identical network visualizations but with four nodes and their corresponding links removed in the right one. 
Their task was to identify the missing nodes in the left visualization. 
This task required participants to fully inspect the information of \emph{two node-link diagrams}
and was used by Kotlarek et al~\cite{Kotlarek20}. 

\added{More specifically, users saw two diagrams placed side-by-side where the rotation, position, and scale (or zoom level) of these two diagrams were synchronized such that manipulating the left diagram would cause the same manipulation of the right diagram.
Same as in the \tCommon{} task, participants clicked on a node to select or unselect it as one of their answers, \added{and clicking the node changed the node color from its original black color to red or from red back to black}. 
Once participants were satisfied with their choices, they clicked a ``Done" button on the page to submit their selections and stop the internal timer.}

\added{Given the density of nodes, we initially believed this task required participants to exert effort zooming in on and rotating the graphs (when applicable) to find missing nodes and links.}
Each study trial had 100 nodes roughly separated into 3 even-sized clusters.
The probability that a link existed between two nodes was 0.12 within a cluster and 0.02 between clusters.

To generate realistic data throughout all of our tasks, we used the stochastic block model~\cite{holland_stochastic_1983} to create network data with clusters, which are ubiquitous in social, biological, and geographical applications.
During data generation, we pre-determined the number of nodes, the number of clusters, and the probability that a link would exist between any two nodes in the graph. 
We conducted multiple internal tests to determine the optimal combination of node and link probabilities, taking into account the cognitive load burden for network visualizations~\cite{Yoghourdjian20} and the benefits of stereo and motion cues on acceptable data size~\cite{Ware05}.

\subsection{Measures}
\label{sec:measures}
We measured \emph{time} from the instance the visualization was fully rendered to when the internal timer would stop. 
We measured \emph{accuracy} by dividing the number of correct responses by the total number of trials.
We measured \textit{workload} by using an adapted version of NASA's Task Load Index~\cite{Nasa} to rate the four conditions in the areas of \emph{mental demand}, \emph{physical demand}, \emph{temporal demand}, \emph{overall effort}, \emph{frustration}, and \emph{perceived personal performance} for each task.  The subjective ratings were recorded on a Likert 7-point scale.
Finally, at the end of the study, participants were asked to \emph{rank} the conditions in terms of overall effectiveness and \added{general} aesthetics.

\subsection{Hypotheses}
\label{sec:hypotheses}
We developed our hypotheses based on our literature survey and our analysis of the four testing environments along the five main dimensions.

\textbf{$H_{2D-3D-Vis}$}: 
By comparing \cTwoDMouse{} and \cThreeDMouse{}, we wanted to confirm the benefits of using 3D for network visualizations. 
These two conditions share the same display and interaction devices. 
The only distinction is the dimension of visualization rendered on the screen.
We expected \cThreeDMouse{} to outperform \cTwoDMouse{}, as rendering network visualizations in 3D may reduce visual clutter to allow users to navigate to targets more effectively.

\textbf{$H_{tangibility}$}: 
By comparing \cThreeDMouse{} and \cThreeDTrackBall{}, we wanted to verify the benefits of tangibility in navigating network visualizations.
The only difference between the two conditions is whether the condition provides tangible interactions.
We expected \cThreeDTrackBall{} to outperform \cThreeDMouse{} because, with a physical proxy, trackball users can more intuitively map their desired 3D rotations in digital space to the real 3D movement of their physical trackball.
Since a standard mouse lacks this functionality, it requires higher context-switching costs.

\textbf{$H_{direct-interaction}$}:
By comparing \cThreeDVR{} to other conditions, we wanted to identify whether there were real benefits to using direct interaction in VR, which was our most embodied condition in our study.
We expected \cThreeDVR{} to outperform other conditions because with stereoscopic vision and 6DOF tracked controllers, users had an identical display and interaction space that could allow them to directly manipulate 3D objects in their view and possibly increase their efficiency.

%% file: body/5-results.tex
\section{Results}

We used \textit{linear mixed-effect modeling}~\cite{Bates2015} on the logarithmic transformation of completion time, which we found to meet the normality assumption. 
Compared to repeated measure ANOVA, linear mixed modeling is capable of modeling more than two levels of independent variables and does not have the constraint of sphericity~\cite{field2012discovering}.
We modeled all independent variables and their interactions as fixed effects. A within-subject design with random intercepts was used for all models. 
We evaluated the significance of the inclusion of an independent variable or interaction terms using a log-likelihood ratio. 
We then performed \emph{Tukey's HSD} post-hoc tests for pairwise comparisons using the least square means~\cite{Lenth2016}. 
We used predicted vs. residual and Q---Q plots to graphically evaluate the homoscedasticity and normality of the Pearson residuals respectively. 
For accuracy, ratings, and rankings that did meet the normality assumption, we used a \emph{Friedman} test to evaluate the effect of the independent variable, as well as a \emph{Wilcoxon-Nemenyi-McDonald-Thompson} test for pairwise comparisons. Significance values are reported for $p < .05 (*)$, $p < .01 (**)$, and $p < .001 (***)$, respectively, abbreviated by the number of stars in parentheses.

\begin{figure}
    \centering
    \includegraphics[width=\linewidth]{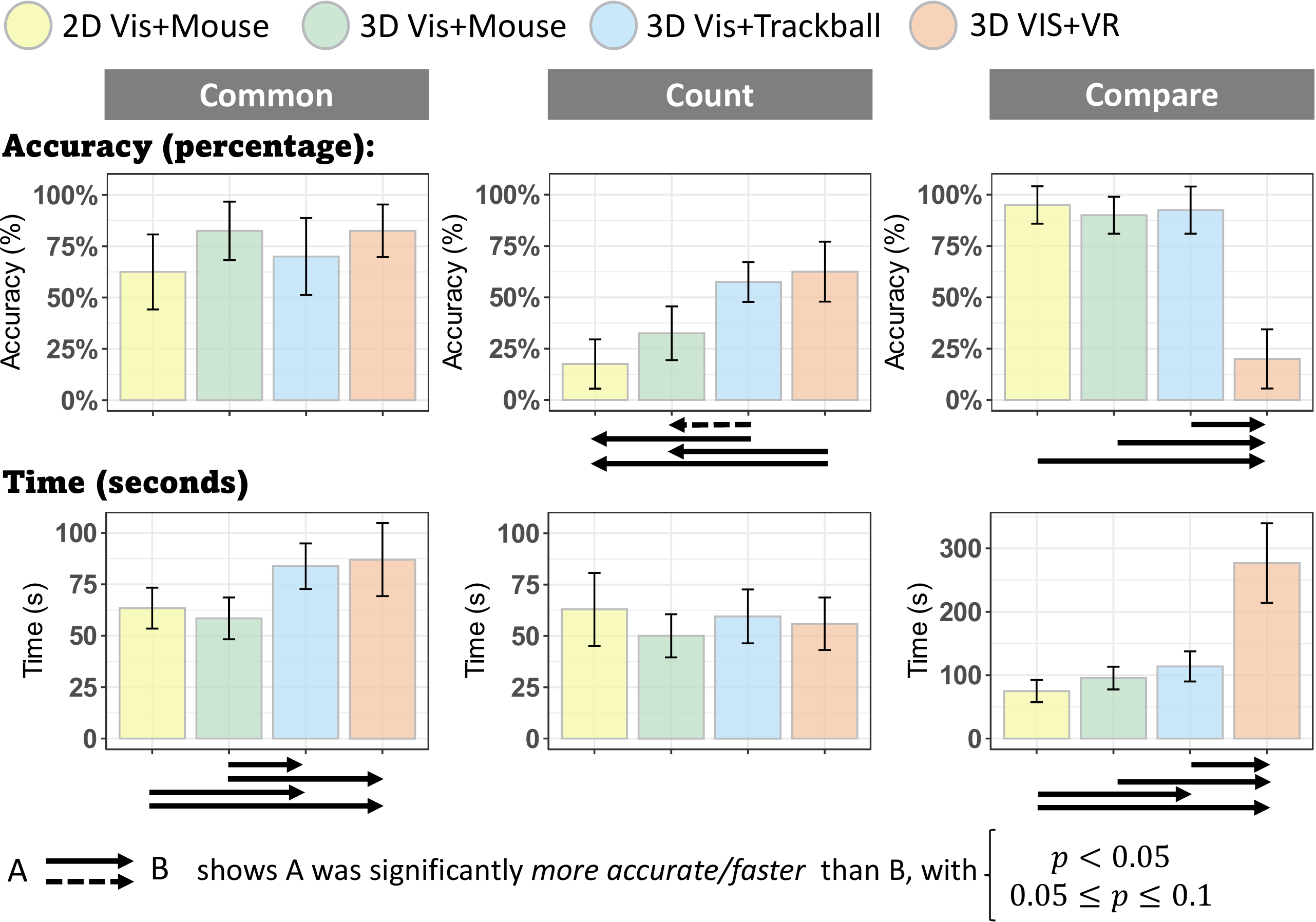}
    \caption{Results for \textit{time} (seconds) and \textit{accuracy} (percent average) by task.
    Error bars are for 95\% confidence intervals.
    }
    \label{fig:taskvalues}
\end{figure}

\begin{figure}
    \centering
    \includegraphics[width=\linewidth]{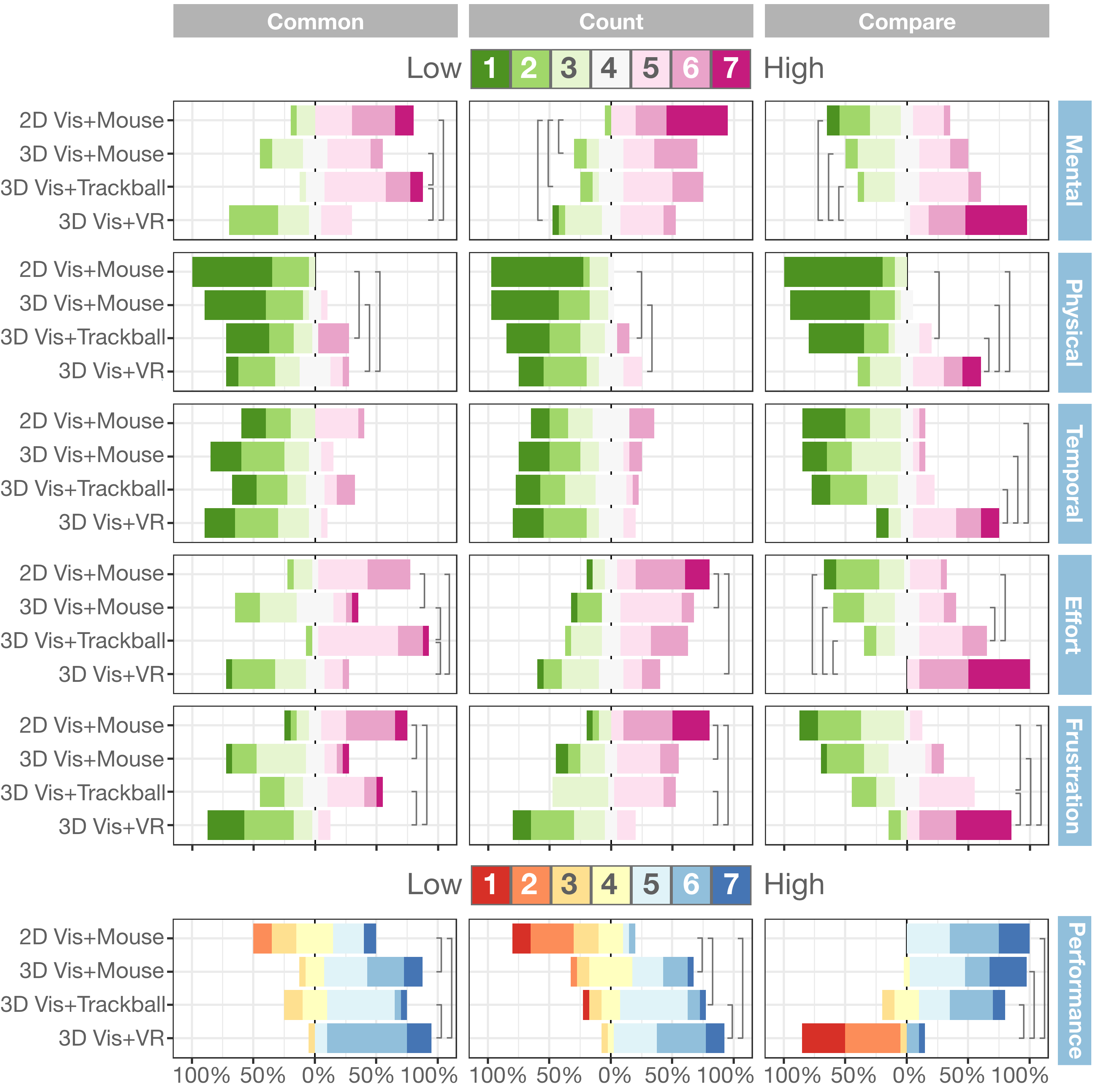}
    \caption{Ratings indicated by participants on a 7-point Likert scale. 
    The bars represent the distribution of scores across all subcategories, and bars closer to the left represent lower workload in the green to pink series, while bars closer to the left represent lower success in the red to blue series. 
    Brackets indicate significances for \textit{p} $<$ 0.05.}
    \label{fig:tlxSummary4}
\end{figure}

\begin{figure}
    \centering
    \includegraphics[width=\linewidth]{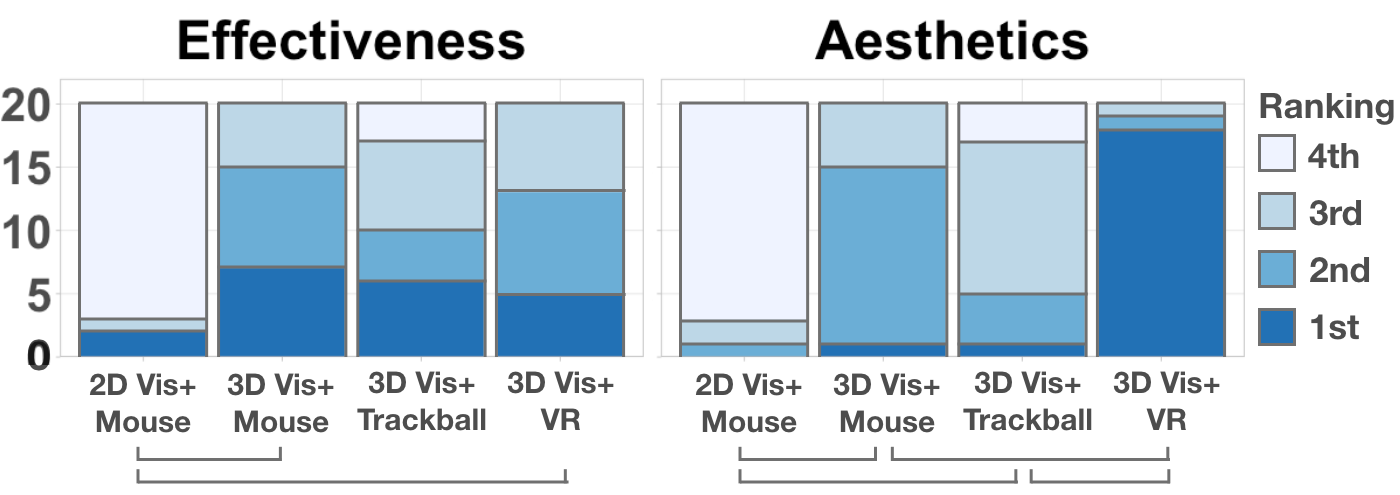}
    \caption{Participant rankings of the four conditions from most to least effective (left) and aesthetic (right) when working with node-link diagrams in general. Brackets indicate significance for \textit{p} $<$ 0.05.}
    \label{fig:postrankings}
\end{figure}

Results for time and accuracy are shown in \autoref{fig:taskvalues}.
\autoref{fig:tlxSummary4} presents participants' task load index responses, and \autoref{fig:postrankings} demonstrates participants' overall effectiveness and aesthetics ranking of the four conditions.
We report on the rejections and acceptances of our hypotheses 
in each task and share the feedback from our participants for each condition.
All detailed statistical results are presented in a supplementary document.

\subsection{2D vs. 3D network visualizations}
We did not find significant differences between \cTwoDMouse{} and \cThreeDMouse{} in time and accuracy for all three tested tasks.
\cThreeDMouse{} (82.5\%, CI=14.2\% in \tCommon{} and 32.5\%, CI=13.1\% in \tCount{})
tended to be more accurate than
\cTwoDMouse{} (62.5\%, CI=18.3\% in \tCommon{} and 17.5\%, CI=12.0\% in \tCount{}), but not by a statistically significant amount.
Since \cTwoDMouse{} and \cThreeDMouse{} had similar performance, we rejected $H_{2D-3D-Vis}$ in terms of time and accuracy.

In terms of workload, participants found that \cTwoDMouse{} required more effort ($*$ in \tCommon{} and $**$ in \tCount{}) and resulted in more frustration ($*$ in \tCommon{} and $**$ in \tCount{}) compared to \cThreeDMouse{} in \tCommon{} and \tCount{}. 
Participants also felt more mental demand in \cTwoDMouse{} than in \cThreeDMouse{} for \tCount{} ($**$). 
Since participants rated \cThreeDMouse{} to be subjectively more effective in some perspectives, we accepted $H_{2D-3D-Vis}$ in task load index ratings.

For overall perceptual rankings, participants ranked \cThreeDMouse{} over \cTwoDMouse{} for effectiveness ($**$) and aesthetics ($***$). Thus, we accepted $H_{2D-3D-Vis}$ in the subjective rankings.

In summary, our quantitative results show a similar performance between \cTwoDMouse{} and \cThreeDMouse{}. However subjectively, participants found that \cThreeDMouse{} required less workload compared to \cTwoDMouse{}.

\subsection{The effect of tangibility}
\added{Again, we defined tangibility as the environment's ability to allow the user to more naturally ``touch" or manipulate the visualization, whether in the form of a physical proxy or a 3D virtual representation.}
We found that \cThreeDTrackBall{} (57.5\%, CI=10.0\%) was marginally more accurate than \cThreeDMouse{} (32.5\%, CI=13.1\%) in \tCount{} ($p=0.1$), \added{but that} \cThreeDMouse{} (50.1s, CI=10.5s) was faster than \cThreeDTrackBall{} (59.5s, 13.1s) in \tCommon{}.
As a result, for these quantitative measures, we only accepted $H_{tangibility}$ in \tCount{} and rejected it for other tasks.

Workload-wise, participants found \cThreeDTrackBall{} to be more mentally demanding ($*$) and requiring of more effort than \cThreeDMouse{} ($**$) in \tCommon{}.
\cThreeDTrackBall{} was also reported to need more effort in \tCompare{} ($**$).
Thus, we rejected $H_{tangibility}$ according to these task load index ratings.

The overall rankings did not reveal any significant difference between \cThreeDTrackBall{} and \cThreeDMouse{}, so we rejected $H_{tangibility}$ in rankings.

In summary, we found that \cThreeDTrackBall{} was more accurate in one task but slower in another task, and it generally required a higher workload compared to \cThreeDMouse{} as reported by users.

\subsection{Is direct interaction beneficial?}
We found that \cThreeDVR{} (62.5\%, CI=14.6\%) was more accurate than \cThreeDMouse{} ($*$, 32.5\%, CI=13.1\%) and \cTwoDMouse{} ($***$, 17.5\%, CI=12.0\%) in \tCount{}.
However, \cThreeDVR{} (87.0s, CI=17.8S) was slower than \cThreeDMouse{} ($**$, 58.5s, CI=10.2s) and \cTwoDMouse{} ($*$, 63.4s, CI=9.9s) in \tCommon{} and significantly slower (276.8s, CI=62.9S) than other conditions in \tCompare{} (all $***$).
Thus, for quantitative measures, we accepted $H_{direct-interaction}$ in \tCount{} but rejected it for the other two tasks.

\cThreeDVR{} performed well in terms of workload in two tasks. We found that participants perceived less mental demand, temporal demand, frustration, and overall effort in \cThreeDVR{} than in \cTwoDMouse{} or \cThreeDTrackBall{} in \tCommon{} (all at least $*$).
\cThreeDVR{} also had benefits in mental demand, effort, frustration, and performance over \cTwoDMouse{}, and on frustration and performance over \cThreeDTrackBall{} in \tCount{} (all at least $*$).
Participants did find that \cThreeDVR{} required more physical effort than \cTwoDMouse{} in \tCommon{} and \cThreeDMouse{} in \tCommon{} and \tCount{}, but this was merited given that the VR HMDs and controllers naturally allowed for more embodied interactions with the networks. 

\cThreeDVR{} did not perform as well with \tCompare{}, where it was ranked the lowest in all workload categories.
So, for task load index ratings, we accepted $H_{direct-interaction}$ for perceived mentally demand, effort, frustration, and performance in \tCommon{} and \tCount{}, and we rejected $H_{direct-interaction}$ for physical demand in these two tasks.
We also reject $H_{direct-interaction}$ for \tCompare{} more generally.

In terms of overall rankings, participants ranked \cThreeDVR{} over \cTwoDMouse{} for effectiveness ($**$).
\cThreeDVR{} was also ranked higher than all other conditions for aesthetics (all $**$), allowing us to accept $H_{direct-interaction}$ in the subjective rankings.

In summary, \cThreeDVR{} was more accurate than other conditions in one task, but slower in another task. \cThreeDVR{} demonstrated a reduced working load in perceived mentally demand, effort, frustration, and performance, but required more physical movement. 
Furthermore, we found \cThreeDVR{} had a similar performance to \cThreeDTrackBall{} in \tCommon{} and \tCompare{}, but subjective ratings revealed a preference for \cThreeDVR{} over \cThreeDTrackBall{} from participants in these tasks.
We also found that \cThreeDVR{} was clearly not suitable for a task such as \tCompare{}.

\subsection{Qualitative Feedback}
We asked participants to give feedback on the pros and cons of each design. We clustered comments into groups for each condition. 
In this subsection, we share representative feedback along with the number of participants who mentioned a similar concept.

\textbf{\cTwoDMouse{}:} In \textit{Common}, participants voiced frustration with occlusion, as it was not always clear whether a link was connected to a certain node or simply being occluded by that node on the way to another node. 
In situations like this, working with a 2D graph decreased trial accuracy because there was no way for users to reorient the graph and validate their answers. 
As a result, users prioritized their speed over their accuracy. 
\cTwoDMouse{} was ranked as one of the worst conditions for \textit{Common}. 

In \textit{Count}, participants found \cTwoDMouse{} to be extremely difficult. 
The 2D quality of the graph prevented them from having a clear understanding and mental picture of the graph's structure. 
One user specifically mentioned \textit{``In 2D, it was harder to see the subgraphs like you can in 3D''}. 
Because these subgraphs represent potentially important relationships among network elements, this feedback suggests that 2D graphs may not be ideal for tasks involving a thorough understanding of a network's structure. 
Another interesting insight from participants was that the data felt inherently 3D in the \textit{Count} task, so seeing it in its 2D form for \cTwoDMouse{} was frustrating and confusing. One user said, \textit{``clearly, this was a 2D image trying to represent something 3D"}, another noted, \textit{``You're trying to imagine what it might be in 3D"}, and yet another admitted that they \textit{``tried to make a 3D figure in [their] head with the 2D figure"}. 
This was an unexpected result because we hypothesized that users would already be accustomed to seeing 2D graphs and therefore counting subgraphs within them. In reality, user feedback and the nearly unanimous fourth place ranking among users show that \cTwoDMouse{} still has serious disadvantages for structure-related tasks.

In \textit{Compare}, it was surprising to see \cTwoDMouse{}'s popularity, as we expected it to be the least favored condition for this task. 
However, unlike in \textit{Count}, users noted that they \textit{``approached this problem as a 2D problem"}. They likened the side-by-side graphs to photos or images they needed to compare, and the most popular strategy was to interact with the graph as little as possible.

\textbf{\cThreeDMouse{}:} In general, \cThreeDMouse{} performed as expected in rankings and feedback compared to the other conditions. In \textit{Common}, the 3D nature of the graph allowed users to rotate it to check their node selections. As a result, occlusion was much less of a problem in all 3D graphs, not just in \cThreeDMouse{}. Almost every participant mentioned that the mouse was already what they were most familiar with, so with the added benefit of 3D graphs, they did not need to think too hard about their answers before submission. With the newer technology (e.g. trackball and VR HMD), they constantly needed to remind themselves of which command performed which function. 

In \textit{Count}, users expressed how much easier it was to complete the task with a 3D object, but a major difference between users was whether they found the mouse or the trackball to be the most intuitive for 3D \added{object/view} manipulation. Many users noted the importance of rotating the graph for this task as they would if they counted the faces of a tangible object. For the participants who found the trackball more intuitive than the mouse, they attributed this perception to the fact that the movements they made with a mouse were 2D movements (movements on a flat, 2D plane), yet the result was a 3D rotation in the graph.

In \textit{Compare}, a couple of users mentioned that having a third dimension was not as beneficial or harmful for the task. Since the predominant strategy was to compare still graphs instead of interacting with them, the ability to rotate the graph in 3D was not always needed. 
Many stated that \cThreeDMouse{} and \cThreeDTrackBall{} were comparable in this task since so little graph interaction was needed.

\textbf{\cThreeDTrackBall{}:} In \textit{Common}, as well as throughout all tasks, many users described the experience of using the trackball as \textit{``fluid"}. For many participants, the smooth rotation of the ball in 3D space and the resulting effects on the digital 3D graph were very intuitive. Users expressed that they benefited from the impression of tangibly feeling the 3D nature of the graph with the trackball. Along those lines, users claimed that \textit{``you actually feel like you're spinning the graph"} and \textit{``it's a good approximation for what you're trying to do on the screen"}. Not all users felt comfortable using the trackball, though, with many attributing their qualms to their complete inexperience with such a device and the fact that they needed to switch between rotation modes and cursor modes on the trackball. Due to their unfamiliarity, many users said that they needed to think actively about every move they made, making the experience less seamless compared to that with a mouse. 

In \textit{Count}, many of the limitations of the trackball were no longer applicable, which accounts for the improvements in time and accuracy compared to other conditions. One of the biggest learning curves with trackball is relearning how to move a cursor. Since users did not need to select any nodes in this task, they could stay in the trackball's rotation mode and only use the cursor to proceed to the next page. This also helped reduce any confusion from needing to switch between rotation and cursor modes several times, as is necessary for the other two tasks. When users could focus on the rotational function of trackballs that make them more immersive than a traditional mouse, feedback on the \cThreeDTrackBall{} condition became more positive. 

In \textit{Compare}, most users felt like \cThreeDMouse{} and \cThreeDTrackBall{} were comparable since they did not use much graph interaction. 

Given that it was a new type of input device for most participants and not quite as exciting as virtual reality, \textit{3D Trackball's} relatively low ranking is in line with expectations.  

\textbf{\cThreeDVR{}:} User feedback for \cThreeDVR{} contained several compelling themes that made this condition extremely popular in users' rankings, with 14 out of 20 users ranking it first in \textit{Common} and 16 out of 20 ranking it first in \textit{Count}.

In \textit{Common}, two key themes throughout the feedback were (1) embodiment and presence, and (2) network tangibility. Embodiment and presence encompass comments about feeling immersed in the virtual space and the data. More than half of the participants mentioned enjoying the ability to move around and specifically move into the graph, which gave them a deeper understanding of the network and its connections. 
\added{As participants completed tasks in VR, we were able to see the virtual environment through their point of view by casting the Oculus Quest 2 headset to a laptop screen. 
Unlike in the other three conditions, in \cThreeDVR{} participants had much more freedom to use head and body movements to complete the tasks, making the screen cast of their actions in VR even more insightful. 
What we observed was that users tended to prefer to stay in place when first entering the virtual environment and use their two controllers to pull the visualization toward them and/or pinch-to-zoom to examine local details.
Only then would they begin to significantly leverage body movements such as moving their head or walking around or "through" the visualization to examine it from slightly different angles.
}
\added{Comments about movement included:}
\begin{itemize}
    \item\added{``You do the rotations of the object, but you can also move yourself without having to rotate the entire graph again. When you rotate a whole object, you have a whole new object to understand. You can just know an object in VR and then move around the [object] you already know."}
    \item\added{``Moving my head was the most intuitive, even more so than using the mouse to rotate things."}
    \item\added{ ``In the other conditions I just wanted to stick my head into the graph and when I got to the VR part, I could actually do that!"}
\end{itemize}
Embodiment ties in with the second theme of network tangibility, which was often stated as a reason for why \cThreeDVR{} felt so intuitive. Many users felt like completing this condition was most like manipulating a real, physical object in \textit{``everyday life"}, and this perception added to the immersive feel of VR.

There were several common downsides to \cThreeDVR{} for this task. One was the precision necessary to point and select a node, which required hand-eye coordination and was difficult for participants with hand tremors. 

In \textit{Count}, similarly to what we saw with \cThreeDTrackBall{}, many of the downsides of \cThreeDVR{} were no longer as applicable because node selection was not necessary. Many of the themes in this task were the same as those in \textit{Common}, but a new theme that stood out was that to some users, 3D graphs only truly seemed 3D in VR, and 3D graphs on a 2D screen still seemed inherently 2D. This task encouraged this observation because many users leveraged head movements to help them quickly and effortlessly count triangles in different parts of the graph. This is not possible with a 3D graph on a 2D screen, where every small change in perspective requires a new click and drag movement. One user did disagree with this idea and said that the VR and 3D conditions were all equivalents, and another expressed that \textit{``[t]he VR gets tiring faster. You get drained faster than if you were using the mouse"}.

Finally, in \textit{Compare}, users strongly disliked \cThreeDVR{}. The most common issue, which nearly every participant shared, was misaligned perspectives. Because the graphs were 3D objects, standing in-between them meant that you were looking at the cluster of nodes and links from slightly different angles when looking at the left and right graphs. The graphs themselves looked sufficiently different from these separate angles. Comparing the two at once proved to be frustrating and difficult, as seen in the \cThreeDVR{} time and accuracy levels in \autoref{fig:taskvalues}.

%% file: body/6-discussion.tex
\section{Discussion}

\textbf{Summary of the results}: which is the winner?
We had mixed results for different tasks.
\textbf{In \tCommon{}}, \cThreeDMouse{} was overall the best performing condition. 
It had a similar time performance to \cTwoDMouse{} and was faster than \cThreeDTrackBall{} and \cThreeDVR{}.
Participants also perceived \cThreeDMouse{} to be more effective than \cTwoDMouse{}.
\textbf{In \tCount{}}, \cThreeDVR{} overall had the best performance.
Its accuracy was similar to \cThreeDTrackBall{}'s and was higher than that of \cTwoDMouse{} and \cThreeDMouse{}.
Subjectively, participants also preferred \cThreeDVR{} over \cThreeDTrackBall{}.
\textbf{In \tCompare{}}, \cTwoDMouse{} demonstrated marginal benefits over other conditions. 
\cThreeDVR{} had the worst performance, while the other three conditions performed similarly in terms of time and accuracy.
\cTwoDMouse{} was rated slightly better than the other two desktop conditions in physical demand, effort, and frustration. 
We discuss the potential reasons for all these findings in the remainder of this section.

\smallskip
\noindent\textbf{Is embodied interaction beneficial?}  
It depends on the nature of the task.
In \tCommon{}, participants were only required to investigate a small portion of the network visualization. 
Being more embodied did not bring significant benefits and in fact likely introduced more overhead.
In \tCount{}, participants had to iterate over the entire graph, which required more intensive interactions and greater predictability of graph manipulation.
This task also required participants to interpret the complex geometry of the network to identify all possible triangles. 
Such a task is better supported by more embodied conditions, as participants can more easily observe the network from different viewing angles. 
This finding about tasks like \tCount{} aligns with a study by Kotlarek et al.~\cite{Kotlarek20}, which found that network visualization in VR performed better than 2D network visualization for interpreting network structure. 
In \tCompare{}, participants were asked to work with two graphs.
For visual comparison, short-term memory plays an important role.
Rendering networks in 3D may result in more information for participants to memorize.
In addition, viewing two items in any natural 3D environment will produce perspective distortion as an individual will look at the items from slightly different angles. This distortion also exists in VR, and participants noted that it seriously affected their performance in \tCompare{}.
In summary, more embodied conditions are better for tasks that are interaction intensive and require the users to frequently manipulate the visualization in order to inspect it from different perspectives.
On the other hand, embodied interaction 
was not ideal for tasks that require memorization in our study.
The anticipated effect of embodied interaction has been reported in many other applications~\cite{hand_survey_1997,besancon_state_2021,ens_grand_2021}, and we now provide extra empirical knowledge about it within the context of navigating network visualizations.

\smallskip
\noindent\textbf{Should we use 3D for network visualization?}  
Previous studies found rendering networks in 3D reduces visual clutter and better facilitates a user's ability to complete visual analytics tasks~\cite{Ware94,Ware05,Ware08,Kwon16}.
We confirmed this finding in \tCommon{} and \tCount{}.
The benefit of 3D was primarily reflected in participants' accuracy scores, which were lower with 2D network visualizations for these two tasks.
We believe that the main reason for lower accuracy with 2D visualizations was the inevitable node/edge overlaps in 2D visual representations, which introduced ambiguities in interpreting network connectivity.
Notably, such a limitation with 2D not only existed in visualizing large network data (as tested in \tCommon{} with 60 nodes) but also in small data (as tested in \tCount{} with only eight nodes).

In addition, 3D network visualizations allowed participants to inspect network connectivity from different viewing angles. The connectivity could be ambiguous from one perspective, but participants had the opportunity to inspect and confirm it from different viewing directions by rotating the graph.
While participants could also zoom in on the overlaps to inspect visual ambiguities in 2D network visualizations, zooming in and out made participants lose track of the context. As a result, they were likely to miss or re-count certain areas of the graph, which could be one of the reasons for their low accuracy in 2D.
We anticipated that the zooming interactions in 2D resulted in higher context-switching costs than rotating in 3D network visualizations.

An alternative way to confirm connectivity within a 2D network visualization is to allow participants to \emph{drag} around the nodes of the graph. However, performing such interaction would likely increase completion time in our tasks and possibly also increase perceived task loads. 
We did not include this alternative method in \cTwoDMouse{} as we wanted to focus on navigation interactions. To achieve this goal, we needed to keep all interactions consistent across conditions (e.g., just pan\&zoom).

Our results demonstrated the limitation of showing detailed network connectivity with 2D network visualizations, and future studies are needed to investigate the effect of providing extra interactions for 2D network visualizations on overcoming such a limitation.
Despite not being efficient in showing detailed network connectivity, 2D network visualization performed better than 3D ones in \tCompare{}.
For this tested task, changing the viewing perspectives in 3D network visualizations introduced a heavier working memory load, as participants needed to re-interpret two visualizations every time after rotating. 
Thus, we believe 2D network visualizations had a better performance because they induced less overhead of this kind.

\textit{In summary,} we found that both 2D and 3D visualizations have their own advantages:
3D network visualizations have the capacity to clearly present detailed network connectivity, and 2D network visualizations are likely to lower the working memory load during detailed visual comparison.

\smallskip
\noindent\textbf{Tangibility vs. familiarity.}  
The trackball mouse is designed to be optimized for ergonomics and accuracy.
When used properly and with familiarity, it allows for smaller, more precise movements compared to the large, rapid movements possible with a traditional mouse. 
In our study, \cThreeDTrackBall{} demonstrated some advantages in accuracy over a traditional mouse in \tCount{}, but the trackball mouse was slower in \tCommon{}.
We believe that participants were more accurate with a trackball mouse because the intuitive interaction with the trackball better allowed them to confirm their answers.

However, the extra interaction introduced more completion time.
Another reason for the longer completion time could have also been due to unfamiliarity with a trackball mouse.
Although many participants commented that the trackball mouse made intuitive sense, they reported high amounts of workload with the trackball.
When designing interactive visualization systems, we should consider the trade-off between the benefits of tangibility and the issues with unfamiliarity when using devices such as a trackball mouse.
Interestingly, while few participants had strong familiarity with VR devices, the broader participant group did not rank \cThreeDVR{} as requiring a high workload as they did with the trackball.
Consequently, we argue that a VR HMD is easier and more intuitive to use than a trackball mouse.

\smallskip
\noindent\textbf{Perspective distortion from stereoscopic vision degrades the performance of certain tasks.}  
We identified two potential reasons for the poor performance with \cThreeDVR{} specifically in \tCompare{}.
First, although the rotation, position, and scale of the two node-link diagrams were synchronized, the VR environment allowed for natural perspective distortion in \tCompare{} that prevented the viewer from effectively comparing the diagrams from the same viewing angle. 
The effect of perspective distortion has been discussed in textbooks~\cite{munzner2014visualization}, and \added{we provide preliminary affirmation in our study.}
\added{While this may seem like a limitation, we believe this finding emphasizes an important aspect and consideration of using virtual reality in certain contexts.} 

Secondly, while large display size is generally considered an advantage, it made \tCompare{} more challenging in VR.
Moving too closely to one graph prevented users from simultaneously seeing the other graph, making comparison difficult. 
Alternatively, scaling (``zooming") in too much on one graph also scaled the other graph, entangling the graphs and inhibiting easy comparison. This kind of entanglement did not happen in our limited desktop display space, where artificial view boxes helped ``crop out'' parts of the graph outside of the boxes.

%% file: body/7-conclusion.tex
\section{Conclusion and Future Work}

We compared four experimental conditions across three network analytic tasks with 20 participants in a controlled user study. 
We found that there is no condition that can always outperform other conditions in all tested tasks.
The more embodied conditions are more suitable for interaction-intensive tasks that require users to constantly manipulate the visualization.
This is because rendering network visualizations in 3D alleviates the issue of visual clutter and improves a user's ability to identify geometric structures.
However, rendering network visualizations in 2D is \added{likely} better for tasks involving graph comparison because it requires less short-term memory usage from users.
Ultimately, our results \added{provide some initial evidence for} empirical knowledge related to the effect of embodiment on navigating network visualization, as well as \added{some} general guidelines for selecting display and interaction devices when working with these graphs.

In future studies, we hope to recruit participants with significant trackball or VR experience to investigate the effects of device unfamiliarity and novelty, as we recognized that our participants were far more familiar with the standard mouse than the trackball mouse or VR HMD.
For example, we will specifically target users in the carpal tunnel syndrome (CTS) community, as its community members are used to trackballs to alleviate wrist pain.
It also should be acknowledged that participants might be sensitive to input settings. Future studies should investigate methodologies to customize the optimal settings for each participant before proceeding to the tasks.
In addition, due to the difficulty of comparing two diagrams in VR, we will potentially use an egocentric design to \added{slant the graphs toward the user in VR and reduce natural perspective distortion}. A similar design has been considered for arranging small multiples~\cite{liu_design_2020}.
We also intend to test different network layout algorithms and more interactions (e.g., highlighting, brushing, and filtering) in the future.

This study is meant as a first assessment of different commercialized devices. 
Although we have identified some benefits of tangible user interfaces for visualization, we only used simple devices that were widely available.
We believe more sophisticated customized tangible devices could be more effective for certain types of tasks, such as the ones in~\cite{Englmeier19,englmeier_tangible_2020,satriadi2022tangible}, and those devices could be interesting to test in future studies.
\added{Additionally, the design space of embodiment on desktop and in VR is huge. We only tested four conditions that we considered as the most accessible in this study. Considering the various input and display modalities (e.g., bare-hand interaction on desktop and mouse interaction in VR~\cite{zhou2022depth}), there is a need to study the spectrum of embodiment in more fine-grained levels for more nuanced insights.}